\pdfoutput=1

\documentclass{jkas}

\def\year{2014} 
\def\month{April} 
\def\volume{47} 
\def\issue{2} 
\def\beginpage{43} 
\def\endpage{55} 
\def\received{September 13, 2013} 
\def\accepted{February 11, 2014} 
\date{Received \received ; accepted \accepted}

\newcommand{\Hmass}{m_{\rm H}}
\newcommand{\mH}{m_{\rm H}}
\newcommand{\kB}{k_{\rm B}}

\def\aap{A\&A}
\def\apj{ApJ}
\def\araa{ARA\&A}
\def\mnras{MNRAS}
\def\apjs{ApJSS}
\def\pasp{PASP}

\title{
EU$N$HA: a New Cosmological Hydro Simulation Code
}

\author[a]{Jihye~Shin}
\author[b,*]{Juhan~Kim}
\author[a,c]{Sungsoo~S.~Kim}
\author[d]{Changbom~Park}

\affil[a]{ School of Space Research, Kyung Hee University, Yongin, Kyungki 446-701, Korea }
\affil[b]{ Center for Advanced Computation, Korea Institute for Advanced Study, 85 Hoegiro, Dongdaemun-gu, Seoul 130-722, Korea }
\affil[c]{ Department of Astronomy and Space Science, Kyung Hee University, Yongin, Kyungki 446-701, Korea }
\affil[d]{ School of Physics, Korea Institute for Advanced Study, 85 Hoegiro, Dongdaemun-gu, Seoul 130-722, Korea }

\def\email{
kjhan@kias.re.kr
}
\affil[*]{ Corresponding author; e-mail:\hspace{-2mm}\texttt{\email}}

\def\corrauthor{
J. Kim
}

\def\runningauthor{
J. Shin et al.
}

\def\runningtitle{
EU$N$HA: Cosmological Hydro Simulation Code
}

\def\keywords{
cosmology: large-scale structure of universe --- galaxies: evolution --- methods: numerical
}

\def\abstracttext{
We have developed a parallel cosmological hydrodynamic simulation code designed for the study of formation and evolution of cosmological structures. The gravitational force is calculated using the TreePM method and the hydrodynamics is implemented based on the smoothed particle hydrodynamics. The initial displacement and velocity of simulation particles are calculated according to second-order linear perturbation theory using the power spectra of dark matter and baryonic matter. The initial background temperature is given by Recfast and the temperature fluctuations at the initial particle position are determined by the adiabatic model. We use a time-limiter scheme over the individual time steps to capture shock-fronts and to ease the time-step tension between the shock and preshock particles. We also include the astrophysical gas processes of radiative heating/cooling, star formation, metal enrichment, and supernova feedback. We have tested the code in several standard cases such as one-dimensional Riemann problems, Kelvin-Helmholtz, and Sedov blast wave instability. Star formation on the galactic disk is investigated to check whether the Schmidt-Kennicutt relation is properly recovered. We also study global star formation histories at different simulation resolutions and compare them with observations.
}

\begin{document}

\jkashead 


\section{Introduction}
Over the last few decades, state-of-the-art
cosmological $N$-body simulation \citep{kim09,kim11,rasera13,angulo12,kuhlen12} has proven to be a powerful tool for study of the Large-scale Structures (LSS) of the Universe.
The observed distribution of galaxies is well reproduced by
the $N$-body dynamics in terms of the two-point correlations \citep{masaki13} or power spectrum of galaxies \citep{tegmark06},
cosmic topology \citep{choi10,choi13}, and the abundance of groups \citep{nurmi13}
or the largest structure in the universe
\citep{park12}. These statistical studies enable us
to refine the current cosmological model by constraining model parameters
at the percent level of accuracy.

Below the galaxy scale, however,
hydrodynamics begin to dominate gravitational forces.
In the potential well of a dark matter halo,
baryonic matter decouples from the surrounding
dark matter inflow forming distinct and complex inner structures such as galactic disks, bulges,
spiral arms,
and so on.
Therefore, it would be indispensable to include the hydrodynamical effects
if one wants to study galaxy formation and evolution.

The Two Degree Field Galaxy Redshift Survey (2dFGRS; \citealt{col01})
and a series of the Sloan Digital Sky Survey (SDSS; \citealt{aih11}),
for example, have enabled us to study physical properties of and
the environmental effect on
individual galaxies.
In the current paradigm of hierarchical clustering,
it is believed that a galaxy is a aggregate of the merging of smaller objects.
Therefore,
the secular evolution of a galaxy
and interaction between galaxies
began to be investigated in great detail
in the cosmological context.
\citet{par09} reported that the morphology of a galaxy is correlated with that of the nearest neighbors.
This correlation maybe reasonably attributed to hydrodynamic interactions between the galaxy and its neighbors.

An observed galaxy is an accumulated sum of various
physical processes over a long period of time.
Galaxies have been shaped not only by gravity and hydrodynamic forces
but also by other astrophysical processes like cooling, heating,
star formation, and supernova feedback.
Baryonic matter is believed to settle down in the deep potential wells of dark matter
and stars form therein, in high density environments. After consuming all available
fuel by nuclear fusion, massive stars at last
explode as supernovae dispersing energy and
metal-enriched material into the interstellar medium.
This explosion may trigger another star formation, and so the baryonic component starts
another life cycle of metal enrichment. Consequently,
the new generation of stars has higher metallicity than before.
Therefore, a galaxy
may have a wide spectrum of stellar populations and metallicities.

There are many cosmological hydrodynamic simulation codes available today \citep[among others]{kra99,fry00,tey02,osh04,wad04,spr05,wet09,spr10}.
Largely, they can be grouped in two categories; Eulerian and Lagrangian codes.
The Eulerian scheme use regular grids to compute the hydrodynamic
interaction between grids. However, the Lagrangian code is based on
the particles carrying the hydrodynamic properties. Mutual interactions
between particles are computed using the SPH (smoothed particle hydrodynamics).

The GOTPM \citep{dub04} code is especially well designed for massive $N$-body
simulations with an efficient use of memory space and fast calculation speeds.
For example, the Horizon-Runs (HRs; \citealt{kim09,kim11}), which were the largest ones at
that time, were performed with $70-350$ billion particles. The GOTPM code adopts non-recursive walks on oct-sibling tree structures which speeds up the code significantly.

On top of the GOTPM code,
we placed the SPH algorithms to properly simulate cosmological structures
down to the galaxy scale the hydrodynamic forces are more dominant than the gravity.
Among the SPH algorithms, we adopt the entropy-conservation scheme \citep{spr05}.
We used
the CLOUDY 90 package \citep{fer98} to arrange a table of the heating and cooling rates
for reference during the simulation run. We set a global value for the reionization epoch
and the heating rated before and after reionization are different. After reionization, Jeans's mass
may increase and small mass objects are prevented from forming due to baryonic pressure resisting
gravitational contraction. We include the self-shielding of gas particles from uniform UV
backgrounds, which may have a larger effect on the formation of the dwarf galaxies in the
early universe \citep{taj98,saw10}.

This paper is organized as follows:
In Section 2, we present the basic equation of motion of the $N$-body particles.
The hydrodynamic equations are listed in Section 3.
Section 4 is devoted to a description of the
individual time stepping adopted in the code
and Section 5 shows how to implement astrophysical processes in the code.
Simulation tests and the results are discussed in Section 6.
We summarize our work in Section 7.

\section{Basic Equations of Motion}
In an expanding background, the comoving distance ($\mathbf{x}$)
 is related to the physical distance ($\mathbf{r}$) with the scale
factor ($a$) as
\begin{eqnarray}
\mathbf{r} &=& {a\over a_m} \mathbf{x},
\end{eqnarray}
where $a_m$ is the maximum value of the scale factor or the scale factor at $z=0$.
In the GOTPM code, we use the following variables
\begin{eqnarray}
\mathbf{x} &=&{\ell_b\over a_m} {\mathbf{x}}_s,\\
m_r &=& \ell_b^3 \left<\rho_0\right>m_s,
\end{eqnarray}
where ${\mathbf{x}}_s$ is the
position of the particle in unit of the mean particle separation
($\ell_b\equiv L_b/N$), $L_b$ is the simulation box size on one side,
$N$ is the number of grids in one direction,
$m_r$ is the physical mass of the particle,
$m_s$ is the particle mass in simulation unit,
and $\left<\rho_0\right>$ is the mean density at the current epoch ($a=a_m$).
Then,
the physical acceleration on a particle can be expressed as
\begin{eqnarray}
\label{n-basic}
\nonumber
{\mathbf{a}}_r &\equiv& {d^2{\mathbf{r}}\over dt^2}= {m_r\ell_b\over a_m} {d\over dt}\left({\dot{a}{\mathbf{x}}_s+a{\dot{\mathbf{x}}_s}}\right) \\
&=& {\ell_b\over a_m} \left[a\dot{a}^2 {d^2{\mathbf{x}}_s\over da^2} + \left({2\dot{a}^2+a\ddot{a}}\right) {d{\mathbf{x}}_s\over da} + \ddot{a}
{\mathbf{x}}_s\right].
\end{eqnarray}
Henceforth, the subscript $r$ and $s$ mean that the quantity is given
in physical units and simulation units, respectively.
In comoving space, equation (\ref{n-basic}) can be rewritten as
\begin{equation}
{d^2{\mathbf{x}}_s\over da^2} + \left({{2\dot{a}^2+a\ddot{a}\over a\dot{a}^2}}\right) {d{\mathbf{x}}_s\over da}
= {a_m\over a\dot{a}^2\ell_b}  \left({{\mathbf{a}}_r}- {\ddot{a}\over a}{\mathbf{r}} \right).
\label{naa}
\end{equation}
The second term in the parenthesis of the right-hand side of the equation
is the acceleration due to the homogeneous background and can be neglected if the universe is isotropic.

The acceleration, ${\mathbf{a}}_r$, can be decomposed into two parts, due to
the gravitational and hydrodynamical force as below,
\begin{eqnarray}
\nonumber
{\mathbf{a}}_s&\equiv&{a_m\over a \dot{a}^2\ell_b}  {\mathbf{a}}_r
={a_m\over a \dot{a}^2\ell_b}  \left( {\mathbf{a}}_r^g + {\mathbf{a}}_r^{\rm gas} \right)\\
&=& I(a) \left[{-{\sum_j m_s^j {{\mathbf{x}}_s^j \over {|{\mathbf{x}}_s^j|}^{3}} +
{{{\mathbf{a}}_r^{j\rm gas}}\over G\Omega_m\rho_0^c\ell_b(1+z)^2}}} \right]
\label{eq_gr}
\end{eqnarray}
where ${\mathbf{a}}_s$ is defined as the acceleration in the simulation unit,
$I(a) \equiv \left({a_m/ a}\right)^3 ({G\Omega_m\rho_0^c/ \dot{a}^2})$,
$\rho_0^c$ is the critical density at $z=0$,
and the summation index is running over all the other particles.
After some calculations, it can be shown that
equation (\ref{eq_gr}) has the following form;
\begin{eqnarray}
{\mathbf{a}}_s = {1\over a^3B(a)}
\left[{-{\sum_j m_s^j {{\mathbf{x}}_s^j \over |{{\mathbf{x}}_s^j}|^{3}} +
C(\ell_b,z) {\mathbf{a}}_r^{\rm gas}}}\right]
\label{force}
\label{det}
\end{eqnarray}
where $C(\ell_b,z) = 8\pi/[3\Omega_m H_0^2\ell_b(1+z)^2]$ and
\begin{equation}
B(a)\equiv
{8\pi\over3} \left({{1\over a} + {1\over\Omega_i} -1+{\Omega_\Lambda\over\Omega_m}{a^2-1\over a_m^3}}\right).
\end{equation}
Here, the initial matter density, $\Omega_i$ is given by
\begin{equation}
 \Omega_i \equiv \left[{1 + {\Omega_\Lambda\over\Omega_m} \left({1\over a_m^3}\right) +
\left({{1\over\Omega_m} -1 - {\Omega_\Lambda\over\Omega_m}}\right)\left({1\over a_m}\right)
}\right]^{-1}.
\end{equation}

In an expanding medium, it is helpful to split the velocity into two parts;
the radial streaming velocity (or the Hubble flow) and the peculiar velocity.
Then, the Hubble flow
between two points of separation, $x^{ij}_s$ is obtained as
\begin{equation}
{\mathbf{v}}_{\rm H} = g_1{\mathbf{x}}^{ij}_s
\end{equation}
where $g_1 = H(z)\ell_b/(1+z)$. The peculiar velocity (${\mathbf{v}}_p$) is obtained from the simulation
velocity (${\mathbf{v}}_s$) as
\begin{equation}
{\mathbf{v}}_p = g_2 {\mathbf{v}}_s
\end{equation}
 where $g_2= ag_1$.

\section{Hydrodynamics}
We adopt the same entropy-conservation scheme of the SPH as used in the Gadget
code \citep{spr05}. But we use a different method to identify the $N$-nearest neighbors.
Rather than using the predict-correct method adopted by the Gadget code,
we apply an improved method, a direct search
with the Oct-Sibling Tree, which is fast and reliable in identifying the neighbors.
We call this cosmological hydrodynamic code the EU$N$HA
(Evolution of the Universe simulated with $N$-body and Hydrodynamic Algorithms).
\subsection{Basic Hydro Equations}
The pressure ($P_r$) and specific internal energy ($u_r$) can be
measured from the temperature ($T$) and density ($\rho_r$) of ideal gas as
\begin{eqnarray}
P_r &=& {\kB T\over \mu \Hmass} \varrho_{r} \\
u_r &=& {1\over \gamma-1} \left({P_r\over \varrho_r}\right),
\end{eqnarray}
where $m_H$ is the mass of hydrogen atom,
$\mu$ is the mean molecular weight, $\kB$ is the Boltzmann constant, and $\gamma$ is the adiabatic index.
In an adiabatic process, the pressure is related to the gas density
as $P_r = A_r \rho_r^\gamma$, where $A_r$ is the entropy.

Now we adopt the following relations of conversion between the physical and simulation units,
\begin{eqnarray}
\varrho_{r} &=& \left<{\rho}_z\right> \varrho_{s},\\
A_r &=& \left<\rho_z\right>^{1-\gamma} A_s,\\
P_r &=& \left<\rho_z\right> P_s,
\end{eqnarray}
where $\left<{\rho}_z\right> = \left<\rho_0\right>(1+z)^3 = \Omega_m\rho_0^c (1+z)^3 $ and
\begin{equation}
A_s = \rho_s^{1-\gamma}\left({\kB T\over \mu\mH}\right).
\end{equation}
It should be noted that
the simulation entropy ($A_s$)
may change with time even when $A_r$ is unchanged with time.

\subsection{Smoothed Particle Hydrodynamics}
The basic equations of the smoothed particle hydrodynamics are summarized as follows;
{\small\setlength\arraycolsep{0.03em}
\begin{eqnarray}
\nonumber
{\mathbf{a}}_r^{i,\rm gas}&=&-\sum_{j=1}^{N_n} m^j_r \times\\
\nonumber
&&\Bigg[f_i\left({P^i_r\over {\varrho^i_{r}}^2}\right) {\nabla_r^i} W^{ij}_r(h_r^i)
+f_j\left({P^j_r\over {\varrho^j_{r}}^2}\right) {\nabla_r^j} W^{ij}_r(h_r^j)\Bigg] \\
\nonumber
&=&-{1+z\over\ell_b}{A^i_s{\varrho^i_{s}}^{\gamma-2}} \sum_{j=1}^{N_n} m^j_s  \times\\
&&\left[f_i{\nabla_s^i} W^{ij}_s(h_s^i)
+ f_j\left({{A^j_s\over A^i_s} {{{\varrho^j_{s}}^{\gamma-2}}\over{\varrho^i_{s}}^{\gamma-2}}}  \right)
{\nabla_s^j} W^{ij}_s(h_s^j) \right],
\label{fgas}
\end{eqnarray}}
where $h$ is smoothing length defined as the distance
to the $N_n$'th nearest neighbor, $W$ is the smoothing kernel, and $f_i$ is defined as
\begin{equation}
f_i \equiv \left( 1 + {1\over 3}
{\partial \ln \rho_i \over \partial\ln h_i} \right)^{-1}.
\end{equation}
In this study, we used the spline kernel discussed in \citet{monaghan92}.

We have adopted the typical artificial viscosity effect to the acceleration as
\begin{eqnarray}
{d{\mathbf{v_r}} \over dt}|_{\rm visc} &=& -\sum_{j=1}^{N_n} m^j_r\Pi^{ij} \nabla_r^i {\bar W}_r^{ij}\\
&=& -{(1+z)^4{\left<\rho_0\right>}\over \ell_b} \sum_{j=1}^{N_n} m_s^j \Pi^{ij} \nabla_s^{i}
{\bar W}_s^{ij}
\end{eqnarray}
where $\Pi^{ij}$ is the viscosity factor
introduced to capture the shock front.
We adopt the form proposed by \citet{mon97}
\begin{equation}
\Pi^{ij} = - {\alpha\over2{\left<\rho_z\right>}}{(c_s^i+c_s^j-3w_r^{ij}) w_r^{ij}\over\varrho_s^{ij}},
\end{equation}
where $w^{ij}_r \equiv g_1\left( \left| x_s^{ij}\right| + a v^{ij}_{\parallel,s}\right)$
and $\alpha$ is viscosity coefficient. We adopt $\alpha=1$ in this study.
The simulation sound speed, $c_s$, is defined as
\begin{equation}
c_s \equiv \sqrt{\partial P_s\over \partial \varrho_s},
\end{equation}
and this is identical to the physical sound speed, $c_r$.
Finally, we get the viscous force as
\begin{equation}
{d{\mathbf{v}} \over dt}|_{\rm visc} =
{\alpha(1+z)\over2 \ell_b} \sum_{j=1}^{N_n} m_s^j {(c_s^i+c_s^j-3w_r^{ij}) w_r^{ij}\over\varrho_s^{ij}} \nabla_s^{i}.
{\bar W}_s^{ij}
\end{equation}

\subsection{Neighbor Findings}
We use the Oct-Sibling Tree (OST) to find the $N$-nearest neighbor gas particles.
The OST has been extensively used in the Tree force calculation
and has proven to be very fast because of the non-recursive nature of the sibling connections
\citep{dub04,kim11} between Tree nodes and particles.
We exploit the OST for identifying the $N$-nearest neighbors (or smoothing length)
with a simple
modification of the original Tree-gravity routine.

The advantage of the tree searching
is that it identifies the $N$-nearest neighbors without any assumption on the
initial trial value \citep{tha00,spr05}.
Although the predict-correct method has been widely used to determine the smoothing length,
there is a tension between successive iterations
when the smoothing length changes significantly,
especially in regions where the number of neighbors dramatically changes with a small change
of the searching length.

\section{Individual Time Step}
\subsection{Subtime step}
In a particular time step ($\Delta t$),
the change in the position of a particle is quite simply
\begin{equation}
{\mathbf{r}}^\prime = {\mathbf{r}} + {\mathbf{v}} \Delta t + {1\over 2}{\mathbf{a}} \Delta t^2,
\end{equation}
where $\mathbf{v}$ is the velocity and
${\mathbf{a}}$ is the acceleration measured at the position of the particle.
This raises the question of determining the time step $\Delta t$ for which the above expression is
a valid approximation. In many cases, it is sufficient to constrain the time step
by $|{\mathbf{r}^\prime}-{\mathbf{r}}| \leq \epsilon$,
which means that the position change should be less than the force resolution ($\epsilon$)
times the step size
of the simulation.
If the velocity of a particle is larger than the acceleration,
it is reasonable to use
\begin{equation}
\label{velo-ind}
\Delta t_v = \epsilon/|{\mathbf{v}}|.
\end{equation}

The time-step size for a given gas particle is determined as \citep{spr05}
\begin{equation}
\label{ind-hydro}
d t_h = C{h_{sml}\over V_{sig}},
\label{sphind}
\end{equation}
where $C$ is the Courant number, $h_{sml}$ is the smoothing length,
and $V_{sig}$ is the maximum signal velocity of the particle.
The signal velocity between particle $i$ and $j$ is defined as
\begin{equation}
\label{ind-hydro2}
V_{sig}^{ij} \equiv c_s^i + c_s^j - 3 \hat{\mathbf{r}}_{ij}\cdot{\mathbf{v}}_{ij}
\end{equation}
where $\hat{\mathbf{r}}_{ij}$ is the normalized mutual displacement vector
and ${\mathbf{v}}_{ij}$ is the relative velocity.
Here, $V_{sig}$ is the maximum value among $V_{sig}^{ij}$'s.
This ensures that for a given time the ``signal distance'' should be less than the smoothing length scale
multiplied by the Courant factor, which is fixed to 0.15 in this paper.
It is important to note that equation (\ref{ind-hydro}) includes not only the sound speed
but also the relative radial speed between particles.

During the simulation run, particles experience various forces
and their velocities change with time. For example, a supernova explosion
may expel nearby gas particles in the radial direction and, therefore, the time-step size
should be reduced to properly capture the supernova shock.
In the original GOTPM code, however, global time stepping was adopted.
In underdense regions, particle positions vary slowly and hence evolution over
larger periods of times maybe computed in fewer iterations. Thus, adopting global
time steps maybe wasteful and lead to bottlenecks in improving simulation performance.
Hence, in the EU$N$HA code we implemented the individual time steps for both
the $N$-body and hydro parts.

\subsection{Merging $N$-body and Hydro Subtime Steps}
For the EU$N$HA code, we adopt the global time step blocking time stepping, whose subtime step is obtained by dividing the global step size
by the integer power of two;
\begin{equation}
dt = {\Delta t\over 2^p}
\end{equation}
where $dt$ and $\Delta t$ are the individual and global time step sizes, respectively, and $p$ is
an integer.
This equation can be changed to $p = {\rm ceil}(\log_2({\Delta t/dt}))$ where ceil() is a
round up function. Hereafter, we call $p$ the subtime step power.
As we are using two kinds of forces (gravitation and hydro forces),
there are two individual step sizes. Dark matter particles have only $N$-body subtime step
(Eq. \ref{velo-ind})
while gas particles have an additional hydro subtime step (Eq. \ref{sphind}).
If a gas particle has different step sizes,
we adopt the smaller value.

However, the individual time step may not properly capture the shock front.
It is because that a particle in a pre-shock region (low $T$ and $C_s$) may
possibly have a significantly larger step size than the shock-passing time scale.
In some extreme cases, the pre-shock particle may not even experience the shock front due to its large size of time step.
In order to avoid this situation and to relax the tension between
neighboring particles which have a large difference in the step size,
we adopt the time step limiter \citep{sai09}, which propagates the step size
into neighbor particles so that the difference
of the subtime step power among neighbor particles should not be larger than two.
For a full description of the method, see \citet{sai09}.

\begin{figure}[t!]\centering
\includegraphics[width=8cm]{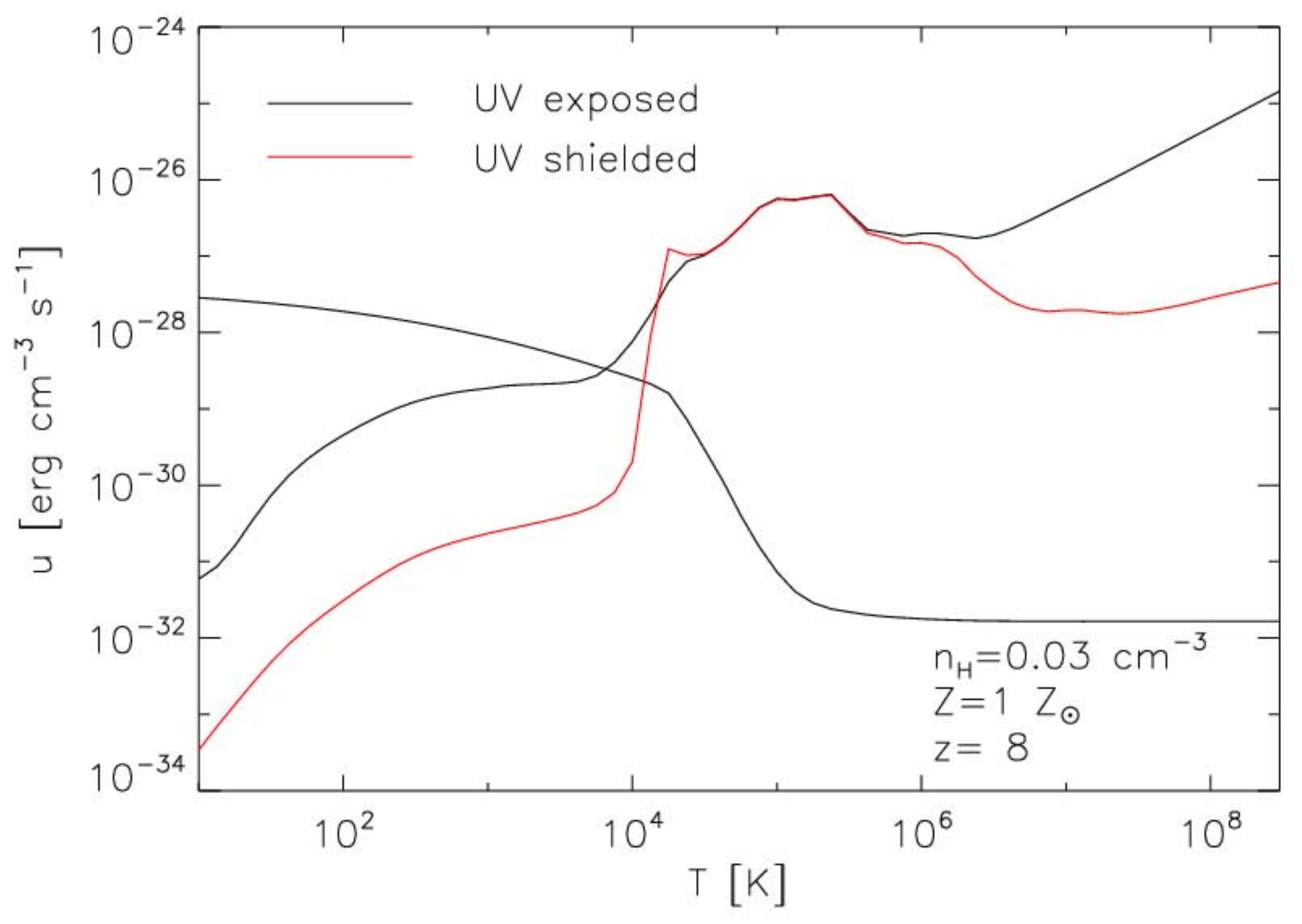}
\caption{Radiative heating/cooling rates of UV exposed/shielded
gas as a function of temperature. The black lines denote the
radiative heating/cooling curves when gas are irradiated by the UV
background radiation, while the red line denotes the collisional
cooling curve where the UV is shielded. Except for the existence
of the UV radiation, the other conditions are the same as
$n_{\mathrm{H}}=0.03$~cm$^{-3}$, Z=1~Z$_{\odot}$, and $z=8$.}
\label{fig01}
\end{figure}

\begin{figure*}[t!]\centering
\includegraphics[width=15cm]{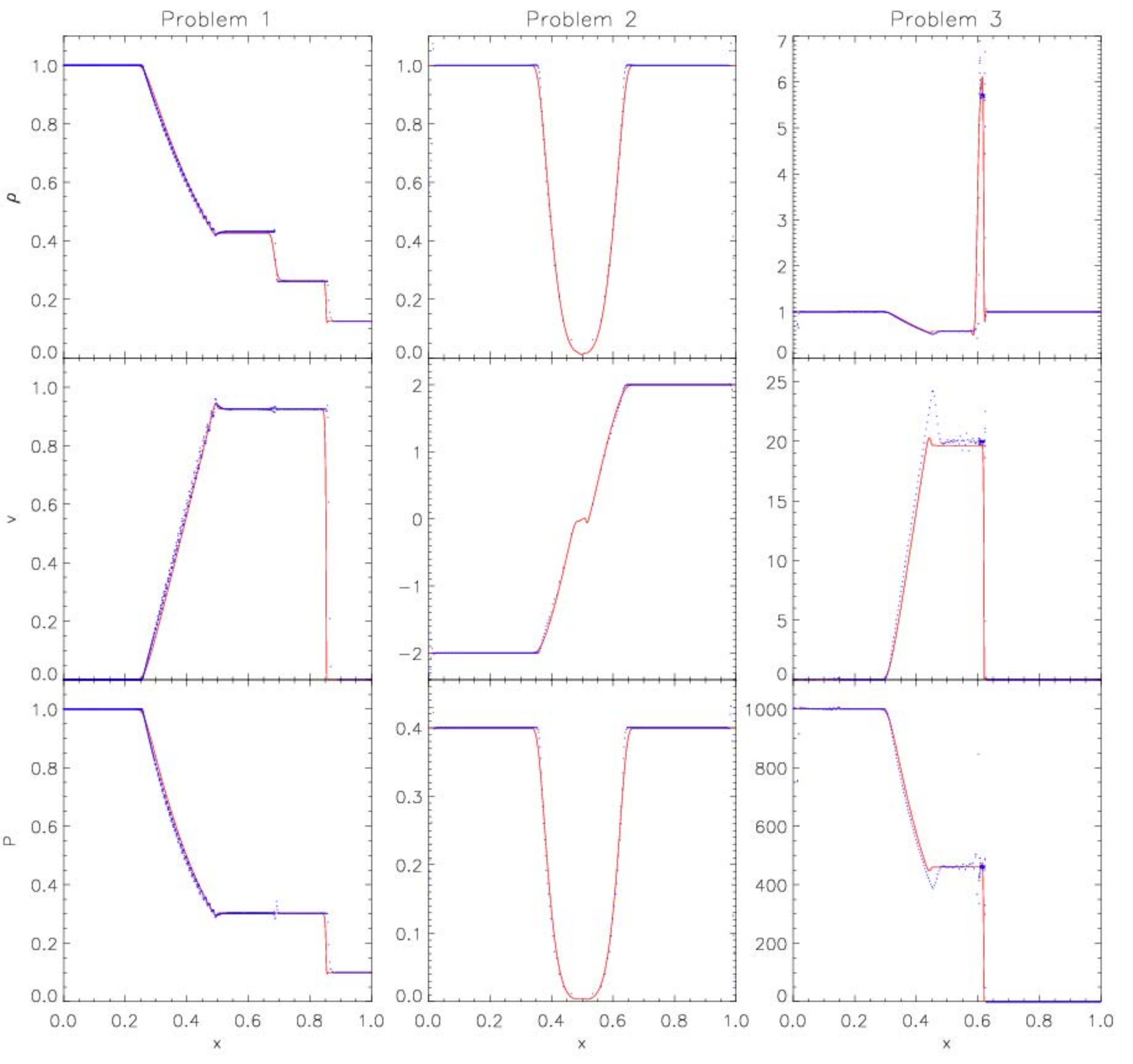}
\caption{Three comparisons of the one-dimensional Riemann shock tube problem
between analytic solution ({\it red line})
and the simulations ({\it blue dots}).}
\label{fig02}
\end{figure*}

\section{Astrophysical Processes}
In addition to the basic hydrodynamic algorithms, we implement
the following astrophysical processes in the EU$N$HA code:
(1) non-adiabatic evolution of the gas particles through radiative
cooling, (2) global reionization heating, (3) star formation, and (4) energy and metallicity
feedbacks by supernova type II (SN$_{\mathrm{II}}$) explosion.

\subsection{Radiative Heating and Cooling}
Using the CLOUDY package (version 10.10; \citealt{fer98}),
we calculated
the heating and cooling rates and tabulated them
in four terms of the gas
density ($\rho$), temperature ($T$), metallicity ($Z$), and redshift
($z$) combining the effects of Compton heating/cooling, inverse Compton cooling,
atomic/molecular cooling, and background UV heating.
We assume that
the whole simulation box is instantaneously full of the UV photons emitted
by massive Pop-III stars at $z_{re}=8.9$ (\citealt{haa96}).
With this step-function like global reionization process, we calculate the collisional
ionization before $z_{re}$ and the photoionization after $z_{re}$.
Also to reflect the self-shielding effect when a dense gas cloud is
optically thick against the background UV radiation, we adopt
the critical hydrogen number density, $n^s_{\mathrm{H}}$, above which the UV
background radiation is effectively blocked. In this study, we set
$n^s_{\rm H} = 0.014~\mathrm{cm}^{-3}$ following
\cite{taj98} and \cite{saw10}. Figure \ref{fig01} shows the heating/cooling
rates of UV exposed/shielded gas. The UV-irradiated gas is gradually
heated up to the equilibrium temperature $T_{\mathrm{eq}}$, at which
the UV heating and molecular cooling are balanced with each other. However, the
UV-shielded gas may continuously be cooled down.

\subsection{Star Formation}
During the simulation run,
we transform a gas particle into a star particle when
the gas particle satisfies
all the following star formation criteria \citep{kat92}:
(1) $n_{\mathrm{H}}>0.1~\mathrm{cm}^{-3}$,
(2) cosmic virialization condition or $\rho_g>57.5\left<\rho_{\mathrm{g}}\right>$, where
$\left<\rho_{\mathrm{g}}\right>$ is the global value of gas density $\rho_{\mathrm{g}}$ at the redshift,
(3) $T<10^4$~K,
and (4) $\nabla\cdot{\mathbf{v}}<0$ for
the convergent flow.

The star formation rate is calculated
according to the Schmidt law as
\begin{equation}
\frac{d\rho_*}{dt}=c_*\frac{\rho_g}{t_{\mathrm{dyn}}},
\end{equation}
where $\rho_*$ is the density of newly born stars, $c_*$ is
the characteristic star formation efficiency, and $t_{\mathrm{dyn}}$
is the dynamical timescale of the gas particle.
In this study, we adopt
the standard star formation efficiency ($c_* =0.0333$) as adopted by
Abadi et al. (2003).
The dynamical time scale of a gas particle is defined as
\begin{equation}
t_{\rm dyn} \equiv \left({{G\rho_g}}\right)^{-1/2}.
\end{equation}
The probability of a gas particle to be a star particle is given by the exponential law as
\begin{equation}
\label{starprob}
P=1-\exp\left(-c_*\frac{dt}{t_{\mathrm{dyn}}}\right).
\end{equation}
and for each time step we determine whether a gas particle satisfying all the
star formation criteria would be converted to a star particle by generating
a random number and applying the probability function in equation (\ref{starprob}).
As the mass resolution of cosmological simulations may not reach
individual stellar mass, each star particle actually represents a star cluster
whose individual stars follow the stellar mass function of \citet{kro01} with the
mass range of 0.1--100~M$_{\odot}$.

\begin{figure*}[t!]\centering
\includegraphics[width=15cm]{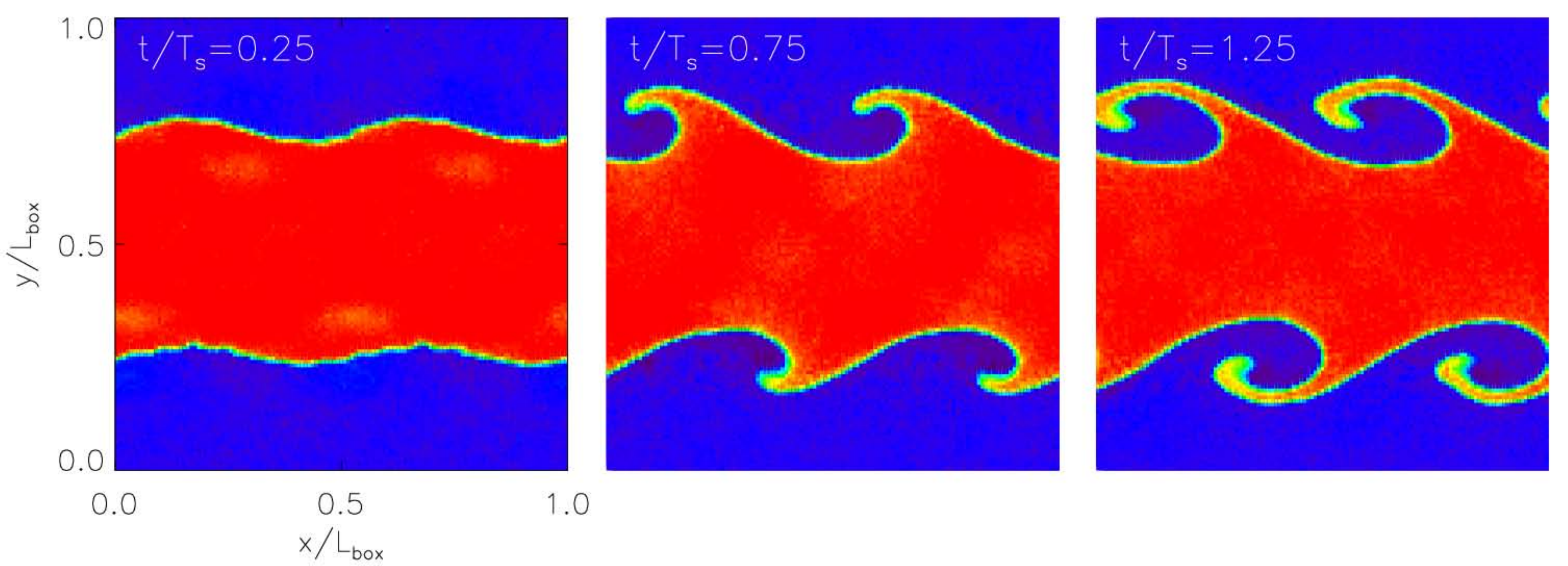}
\caption{Density maps of flows at $t=0.25$, 0.75, and $1.25~T_s$.
The red color denotes high density region moving right,
while the blue color denote low density region moving to the left.
The vortex-like structures are getting prominent with time.}
\label{fig03}
\end{figure*}

\subsection{Supernova Feedback}
Stars follow different evolutionary tracks for different stellar mass
and metallicity. As the  SN$_{\mathrm{II}}$ explosion releases a large amount of energy
and metals, and affects the phase state of the interstellar medium,
it would be important to include the SN$_{\mathrm{II}}$ explosions in a
high resolution cosmological simulation.
For each star particle, we calculate the number of
massive stars which may eventually end up as SN$_{\mathrm{II}}$
explosions using the stellar evolutionary model of \citet{hur00}.
Similar to the star formation recipe, we consider the SN$_{\mathrm{II}}$
feedback using the probability distribution \citep{oka08} as
\begin{equation}
P_{\mathrm{SN}}(Z) = \frac{\int_{t}^{t+dt}r_{\mathrm{SN}}(t',Z)dt'}
                       {\int_{t}^{t_{\mathrm{max}(Z)}}r_{\mathrm{SN}}(t',Z)dt'},
\end{equation}
where $r_{\mathrm{SN}}(t,Z)$ is the SN$_{\mathrm{II}}$ explosion rate
as a function of time and the metallicity ($Z$).
$t_{\mathrm{max}}(Z)$ is the maximum life time of the massive stars.
Typically, $t_{\rm max}(Z)$ is known to be less than ten million years.

We assume that the supernova explosion affects nearby gas particles
through heating and metal enrichment. For this purpose
the SPH-like scheme is adopted to distribute the energy and metals
to the nearby gas particles.
The total amount of energy released by a supernova explosion
is fixed to $10^{51}$ erg.

However the cooling time scale of the shock-heated gas
is usually less than the hydro subtime step.
Then, even though there is a temperature balance between
radiative cooling and background heating, the simulated gas
would not attain the equilibrium. This is the time-resolution problem
in the gas cooling.
To overcome this resolution
problem, we precalculated the temperature evolutions with much finer time steps
as
\begin{equation}
T(t+dt) = T(t) + \int_0^{dt} \left(n_H^2\Gamma(\rho,T,Z)+\Lambda(\rho,T)\right) dt_T,
\end{equation}
where $\Gamma$ is the cooling rate, $\Lambda$ is the heating rate,
and the temperature time step size is numerically set as $dt_T = dt/100$.
At every global time step, we tabulate
the temperature evolution as a function of the gas density, temperature, metallicity,
and time interval ($dt$).
The metal enrichment is measured based on the tables given in \citet{woo95}.

\section{Code Tests}

To test the SPH algorithms and the implemented astrophysical
processes, we perform five test simulations which are the most prominent topics that have been
studied : (1) one-dimensional
Riemann problems, (2) Kelvin-Helmholtz instability, (3) three-dimensional blast shock wave, (4)
star formation on the isolated galactic disk, and (5) global star formation
history in the cosmological context. Since the first three tests are
designed to verify the hydrodynamics, we turn off gravity and
astrophysical processes. We assume static backgrounds except in the last case.
In the last two tests, we include gravity and astrophysical
processes. Cosmic expansion is included only in the last case.

\subsection{One-Dimensional Riemann Problems}

In this subsection, we consider three cases of the one-dimensional Riemann problem
in various shock conditions:
a weak shock (Problem 1), a strong rarefaction shock (Problem 2), and an extremely strong shock (Problem 3).
We adopt the initial conditions given in \citet{spr10} and briefly describe them below.
Periodic boundary conditions are imposed with $0\leq x<1$. The initial conditions simulating a shock
front are chosen in the form of discontinuity in density ($\rho$), pressure ($P$), and
particle velocity ($v$).
Here, the subscript $L$ denotes the region 1 ($x\leq 0.5$) and
subscript $R$ is used for region 2 ($x > 0.5$).
For Problem 1, the initial conditions are
$(\rho,P,v)=(1,1,0)_L$ and $(0.125,0.1,0)_R$ for Region 1 and 2, respectively.
For Problem 2, we set $(1,0.4,-2)_L$ and $(1,0.4,2)_R$.
And for Problem 3, we arrange $(1,1000,0)_L$ and $(1,0.01,0)_R$.
We used $\gamma=1.4$ in these tests
and the number of neighbors is fixed to 7.

We compare the SPH results with the analytical solutions of \citet{sod78} at $t=0.05$ in Figure \ref{fig02}. The density, particle velocity, and pressure are shown for Problem 1 ({left column}), Problem 2 ({middle}), and Problem 3({right}), respectively. In the weak and strong rarefaction shock model, the simulated profiles well match the analytic solutions. However, there are scatters at the contact discontinuity and the shock front is not sharp due to the smoothing features of the SPH. Also, in the extremely-strong shock model we observe that the simulated profiles of the density, velocity, and pressure show larger deviations from the analytic solutions between the rarefaction and the shock front. However, these features are also seen in the standard SPH simulation as shown in other papers.

\subsection{Kelvin-Helmholtz Instability}

\citet{age07} have argued that the standard SPH algorithm cannot
correctly model contact discontinuities like the Kelvin-Helmholtz (KH)
instability. There have been several attempts
to properly handle the instability by modifying the SPH algorithms
\citep{pri08,wad08,rea10}. We examine the
issue of contact discontinuities in the case of the shear flows.

The initial condition of the test is laid out to satisfy the periodic boundary conditions
of the cubic box on a side length of $L_{\mathrm{box}}=0.001~\mathrm{Mpc}$, in which we
distribute $128^3$ gas
particles.
We divide the simulation box into two regions,
the central region (hereafter Region 1) of $\mid y/L_{\mathrm{box}}-0.5\mid<0.25$ and
the outer region (Region 2).
Region 1 has a density ratio of $\rho/\rho_c=2$, where $\rho_c=7.22\times10^7~\mathrm{M}
_{\odot}/\mathrm{kpc}^3$, an $x$-directional velocity of $v_x=
0.5~V_{s}$, where $V_s=8.8~\mathrm{km/s}$. The outer region is set
to have $\rho/\rho_c=1$ and $v_x=-0.5V_{s}$.
Region 1 has a temperature
50000 K and the temperature of Region 2 is doubled
to maintain pressure equilibrium at the contact discontinuity.

We perturb the equilibrium by adding a $y$-directional velocity $v_y$ as
\begin{eqnarray}
\nonumber
v_y(X,Y) &=&  \omega_0\sin(4\pi X) V_s \\
&\times& \left[ e^{-{(Y-0.25)^2/2\sigma^2}}+e^{-{(Y-0.75)^2/2\sigma^2}} \right],
\end{eqnarray}
where $X\equiv x/L_{\mathrm{box}}$, $Y\equiv y/L_{\mathrm{box}}$, $\omega_0=0.1$,
and $\sigma=0.05/\sqrt{2}$ following \citet{spr10}.

Figure \ref{fig03} shows the temporal evolution of density field at
three different epochs, $t=0.25$, $0.75$ and $1.25~T_s$, where $T_s=55.52$
~Myr. As the vertical perturbation grows, the horizontal ($x$-direction) shear
flows generate vortex structures around the contact plane. The expected
whirlpool-like structure gets prominent with time.
It is well known that the standard SPH algorithm has a problem in reproducing
the shear vortex of the KH instability \citep{mcnally12,hubber13,rea10}.
Many authors have suggested that a modification to the SPH algorithms
would be required to produce KH vortex,
for example, the artificial thermal conductivity \citep{pri08,age07}, a suitable smoothing kernel
\citep{valcke10}, the Gudunov-SPH formalism \citep{cha10}, a diffusion term \citep{wad08},
the pressure-entropy formulation \citep{hopkins13}, or the moving mesh \citep{hess10}.
Our findings are similar to that of \cite{pri08} who argued that particle noise
at the discontinuity may suppress the growth of instability.
The typical method to identify neighbor particles is to use the predict-correct algorithm,
which is sometimes erroneous leading to noise.
However, it is too far-fetched to tell
whether our implementation of the neighbor searching method would solve the KH instability problem.
The only thing we can say is that our neighbor finding may be one of the ``partial''
solutions to the KH instability problem.
Further tests would be
required to quantitatively compare EU$N$HA results with others.

\subsection{Three-Dimensional Blast Wave}

\begin{figure}[t!]\centering
\includegraphics[width=8cm]{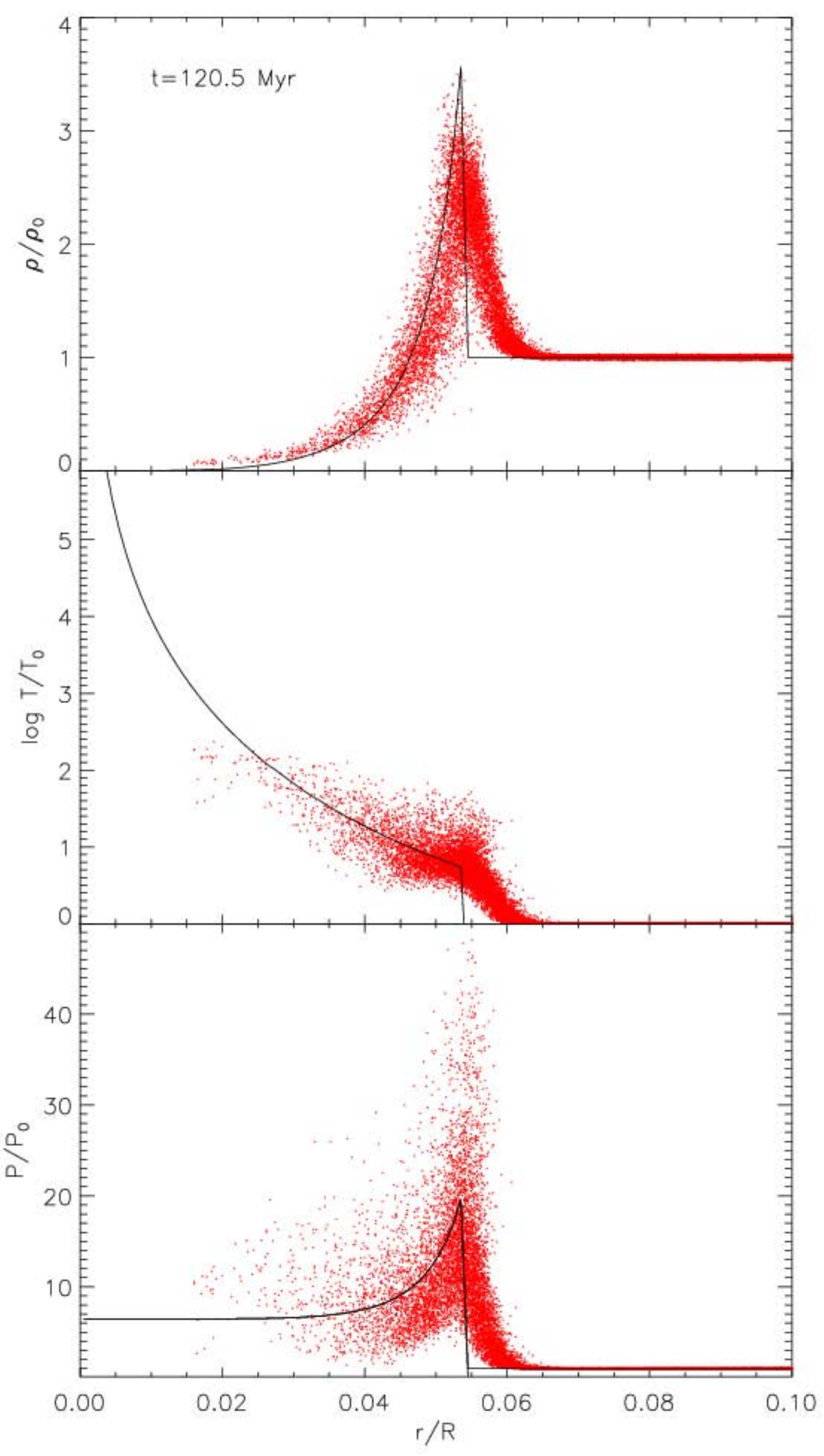}
\caption{Density (top), temperature (middle) and pressure (bottom panel) profiles of three-dimensional Sedov blast wave.
Here, $r$ is the distance from the central particle.
The red dots denote the simulation results and the black line
shows the semi-analytic solution of \citet{sed59}. The $x$-axis is
scaled to the box size and the y-axis is normalized to the initial density,
temperature and pressure of the test simulation.}
\label{fig04}
\end{figure}

We also investigated the three-dimensional
Sedov blast wave test. A nearly homogeneous glass-like distribution of $256^3$
particles is created in the 100~kpc simulation box with total particle mass
$7.22\times10^7$M$_{\odot}$. The temperature of a gas
particle is uniformly set to 10 K.
We assign a huge temperature $T=10^6$~K to the particle located at the center of the box
to simulate the supernova explosion.
This generates a shock wave,
which propagates outwards and sweeps away the surrounding cool and less-dense gas.

Figure \ref{fig04} shows the comparison of density, temperature, and pressure profiles
around the ground zero of explosion between the simulation (points) and
 semi-analytic solution (line; \citealt{sed59}) at $t=120.5$~Myr.
Good agreement is seen both in the location of the shock front and the maximum density,
with a substantial scatter around the analytic solution.
The simulated upstream has a scatter of density with a finite slope
because of the finite smoothing scale of the SPH and the anisotropic initial particle
distribution.
The regular lattice point of initial particle positions was adopted by \cite{spi02} and \cite{mer10}
while a new setup is adopted to reduce anisotropic and inhomogeneous distribution
in the shock propagation \citep{ros07}.

\subsection{Isolated Galaxy}

\begin{figure}[t!]\centering
\includegraphics[width=8cm]{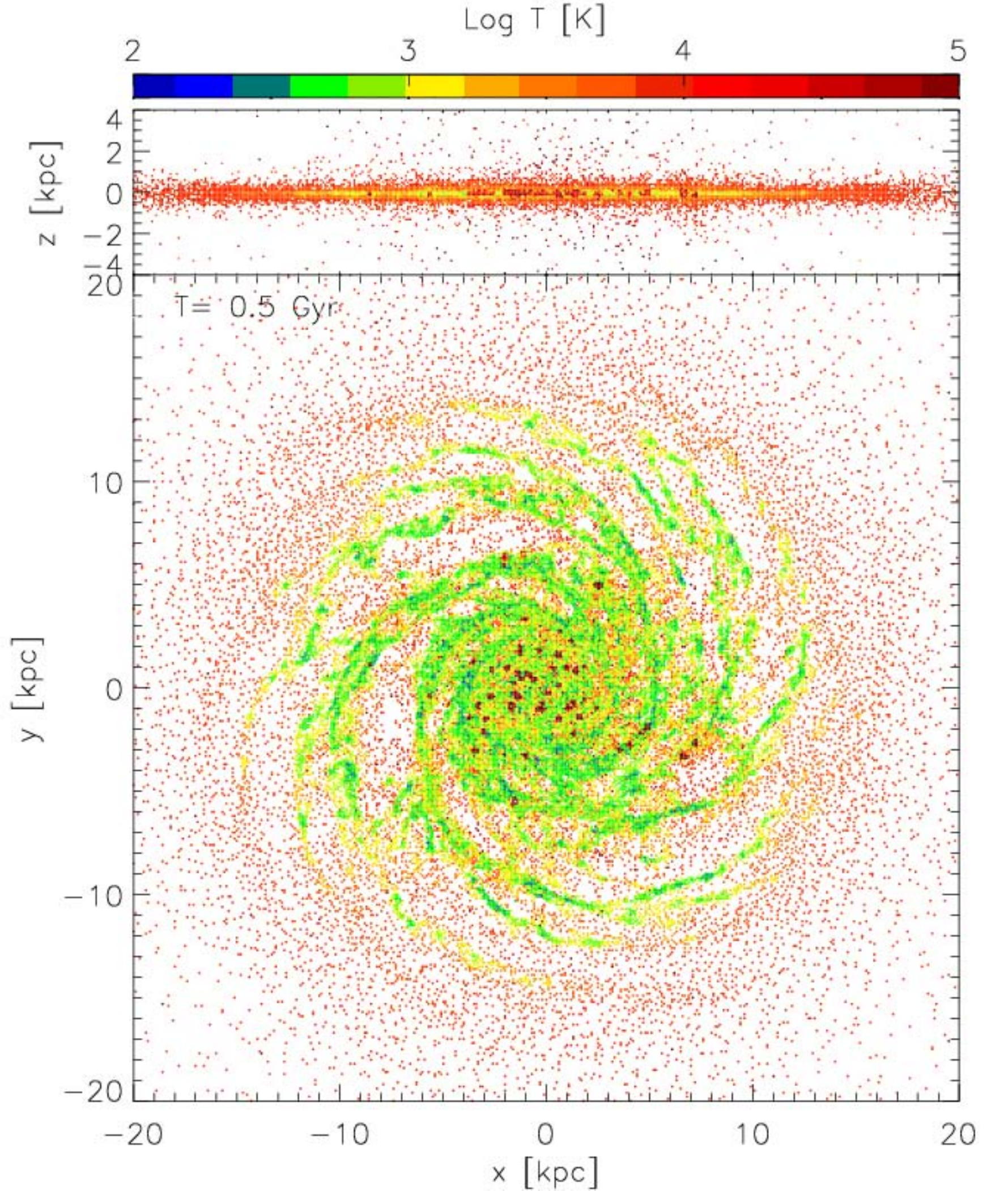}
\caption{Density-weighted temperature map of gas disk. The dense spiral
arms are cooled down more efficiently than the less-dense regions
through the more efficient radiative cooling. Red clumps along the
spiral structures represent the shock-heated gas by the SN$_{\mathrm
{II}}$ explosions among the recently-forming massive stars.}
\label{fig05}
\end{figure}

\begin{figure}[t!]\centering
\includegraphics[width=8cm]{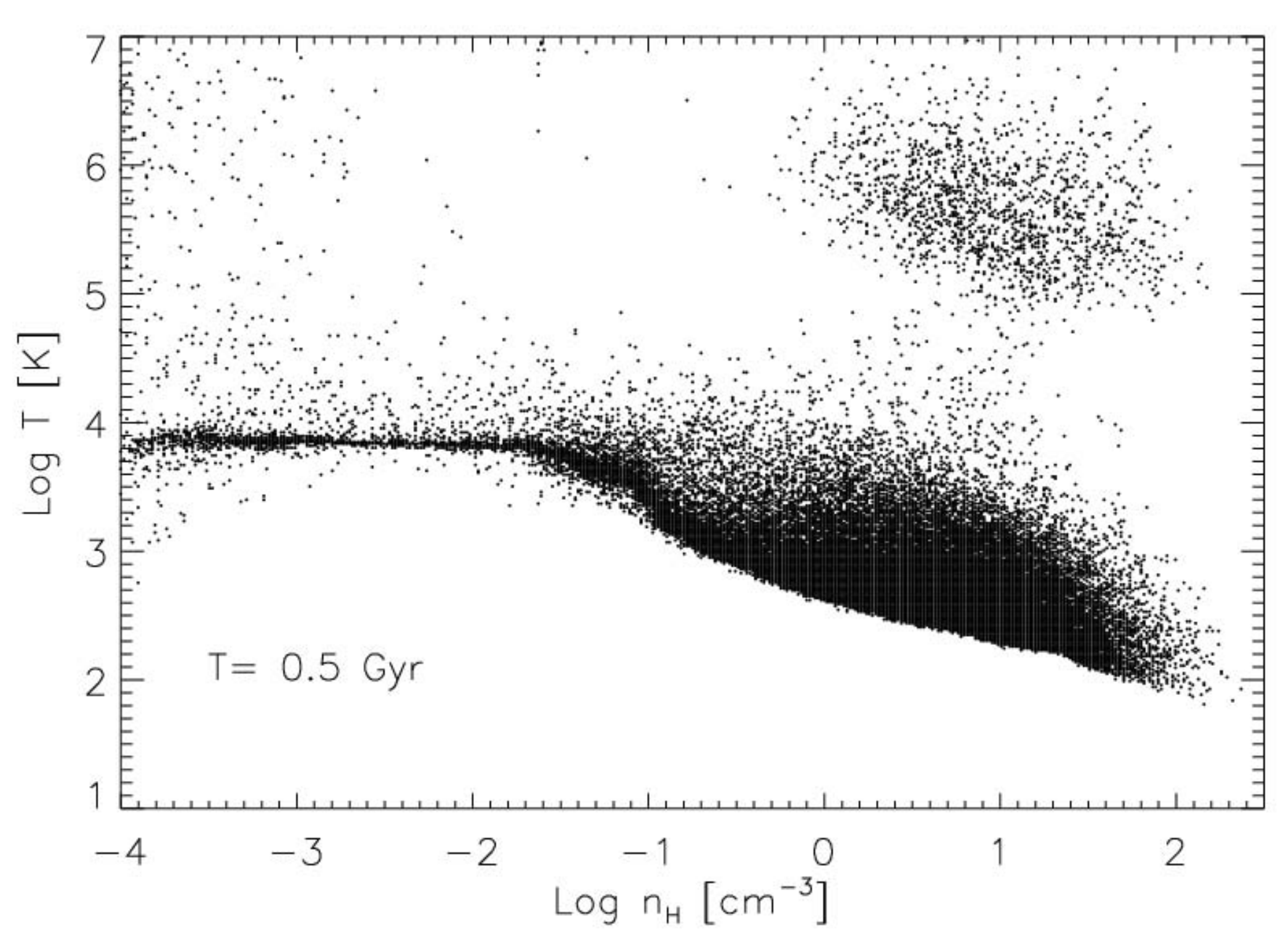}
\caption{$n_{\mathrm{H}}$--$T$ diagram of gas particles at $T=0.5$~Gyr.
All gas particles are irradiated by UV backgrounds because we assume
that there is no self-shielding in this test. Thus,
most of simulated particles are distributed along the equilibrium temperature line.
The scatters  around the equilibrium temperature is mainly
due to adiabatic expansion and contraction. Points in the upper-right corner
are the shock-heated gas particles due to the nearby SN$_{\rm II}$ explosions.
After a transient period, their density will drop due to expansion and points
will move to the upper-left corner of the box.
}
\label{fig06}
\end{figure}
Now, we investigate the stability of galaxy disk
and the star formation rate on the disk plane using the isolated galaxy model.
The initial conditions of a multi-component galaxy were generated according to
the isolated Milky-Way model proposed by \citet{her93}. An exponential
gas disk is arranged with a characteristic mass $M_g=4.125\times10^9\mathrm{M}
_{\odot}$, disk scale height $z=0.3$~kpc, and disk scale length of
$h=3.33$~kpc. An exponential stellar disk is built with $M_s=3.715\times10^{10}
~\mathrm{M}_{\odot}$, $z=0.3$~kpc, and $h=3.33$~kpc. The bulge component
follows Hernquist profile with $M_b=1.375\times10^{10}~\mathrm{M}_{\odot}$ and
scale length of $a=0.8$ kpc. The halo component follows the Hernquist model with
parameters $M_h=2.2\times10^{11}~\mathrm{M}_{\odot}$ and $a=10$ kpc. The
three-dimensional locations and velocities of particles are calculated by
ZENO software package developed by Joshua E. Barnes{\footnote{http://
www.ifa.hawaii.edu/$\sim$barnes/zeno/index.html}}. The gas and stellar
disks are composed of 983,040 particles with particle mass of $M_p = 4.196\times10^4
~\mathrm{M}_{\odot}$ while the bulge and halo structures are treated
as fixed external potentials. We set the initial temperature of
gas particles to be $10^4$~K.

Figure \ref{fig05} shows the density-weighted temperature map of the
gas disk at $t=0.5$~Gyr. Gas particles of $T<10^4$~K are distributed
along spiral structures, where density is so high that the gas
is able to cool
down to lower temperature through efficient radiative cooling.
Density-temperature distribution of gas particles under radiative
heating and cooling process are shown in Figure \ref{fig06}. We can see that
gas particles in the low density regions keep the initial temperature
of $T\sim10^4$~K, while the gas particles in the higher density regions may
cool down to lower temperature toward the equilibrium state.
Stars form in the high-density and low-temperature regions along the
spirals. Tens of million years after star formation,
the SN$_{\mathrm{II}}$ explosions of the massive
stars redistribute
thermal energy into the surrounding interstellar medium. The red clumps
in Figure \ref{fig05} represent the highly heated gas particles
near the supernova explosion sites.

\begin{figure}[t]\centering
\includegraphics[width=8cm]{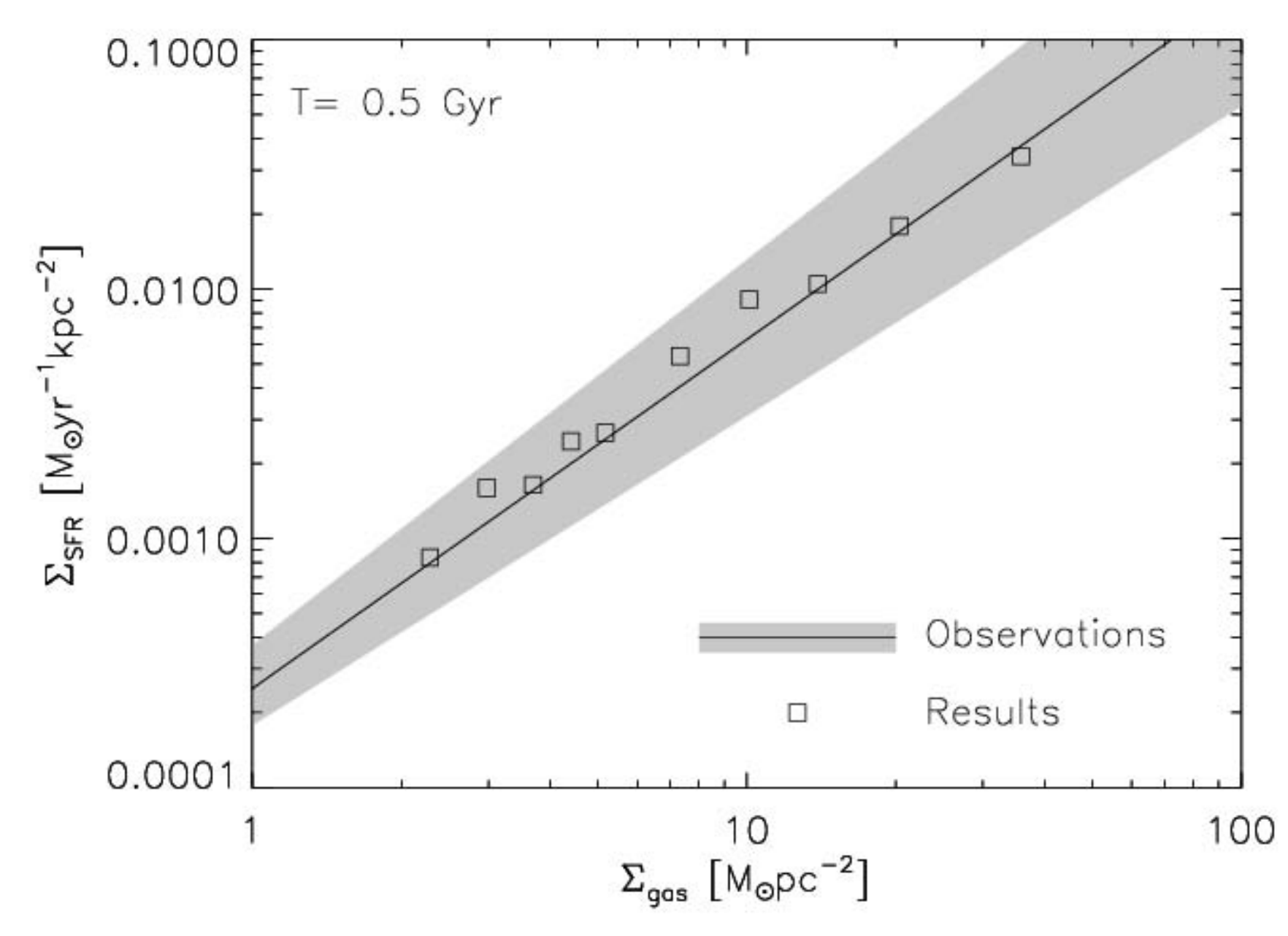}
\caption{Projected star formation rate $\Sigma_{\mathrm{SFR}}$ versus the
column density $\Sigma_{\mathrm{gas}}$ on the simulated disk
at 0.5~Gyr. The solid line enclosed by
the gray-shaded region
represents the observed Kennicutt relation with errors
(Kennicutt 1998), and squares are the simulation results.}
\label{fig07}
\end{figure}

Finally, we compare the simulated star formation rate
with the observed value in Figure \ref{fig07}. The projected star
formation rates on the gas disk $\Sigma_{\mathrm{SFR}}$ for a given column
density $\Sigma_{\mathrm{gas}}$ well reproduce the observed Schmidt-Kennicutt
relation \citep{ken98}.

\subsection{Cosmological Model}

\begin{figure}[t]\centering
\includegraphics[width=8cm]{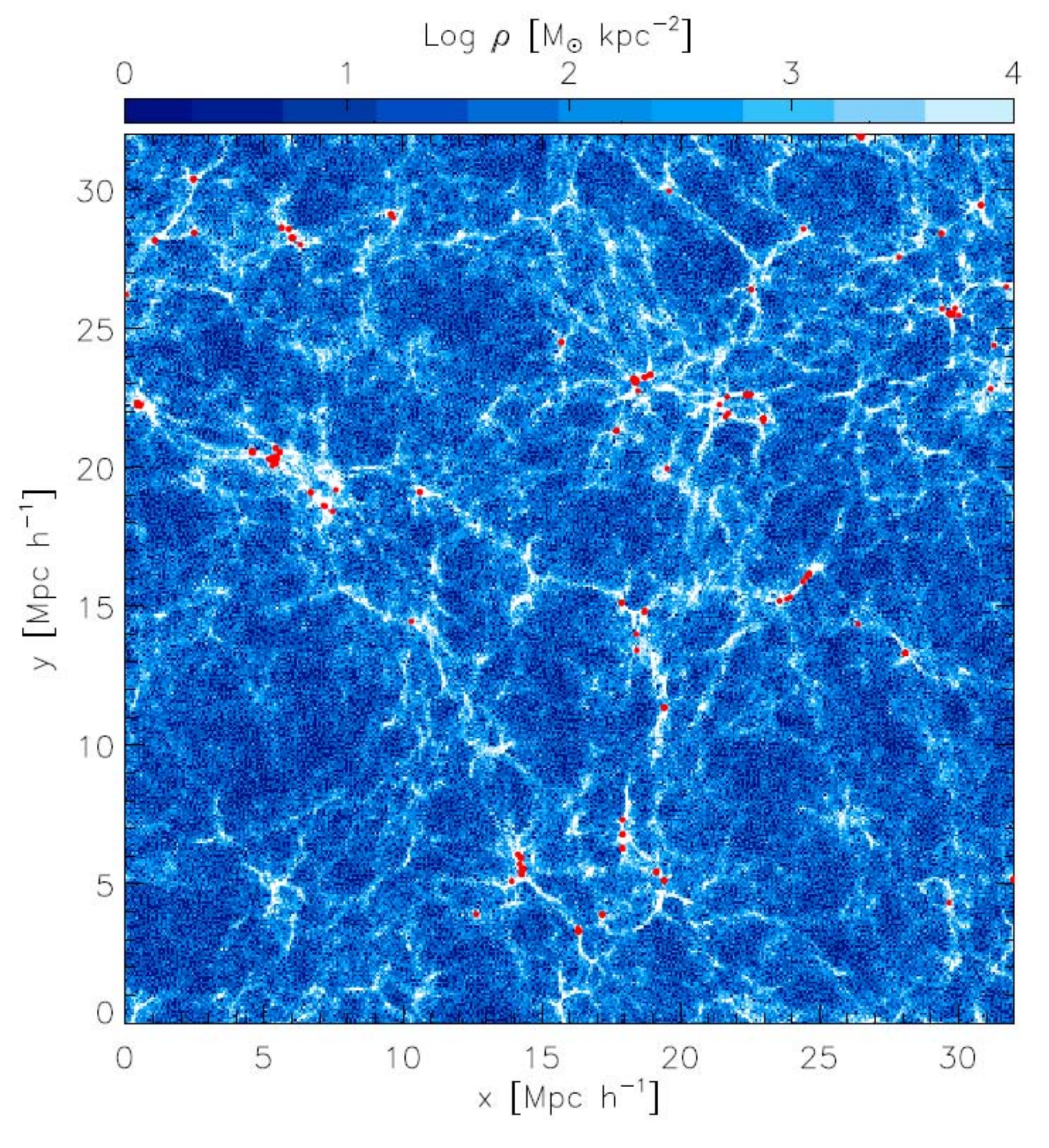}
\caption{Projected density map of gas particles in a $z$-directional
slab with a width equal to three times of the mean particle
separation at the epoch of $z=6$. Red dots represent newly formed star
particles, and they are located in cluster regions.}
\label{fig08}
\end{figure}

\begin{figure}[t]\centering
\includegraphics[width=8cm]{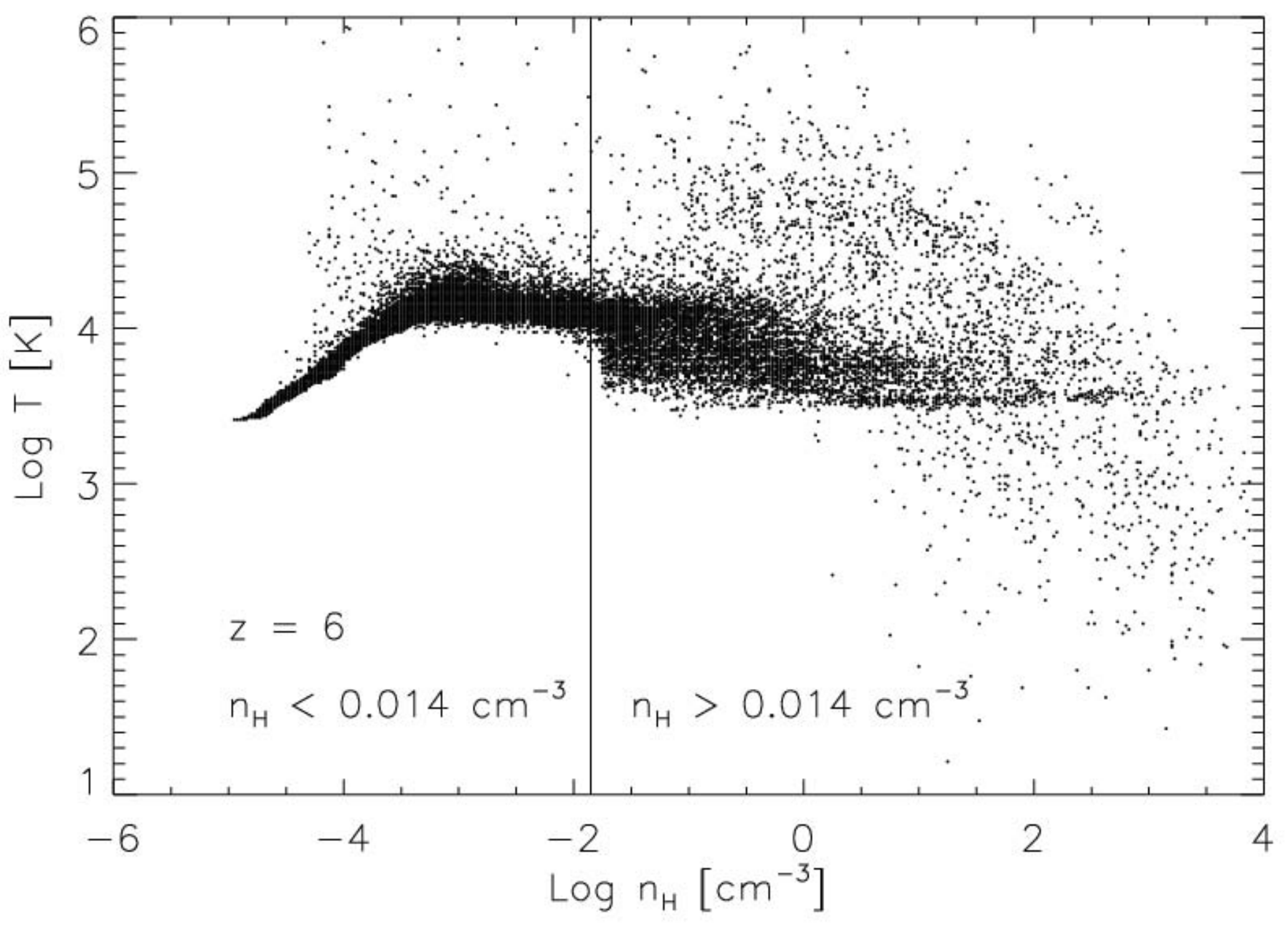}
\caption{$n_{\mathrm{H}}$--$T$ diagram of gas particles at $z=6$.
The solid line marks the UV-background shielding density of
$n_{\mathrm{H}}=0.014~\mathrm{cm}^{-3}$.
}
\label{fig09}
\end{figure}

\begin{figure}[t]\centering
\includegraphics[width=8cm]{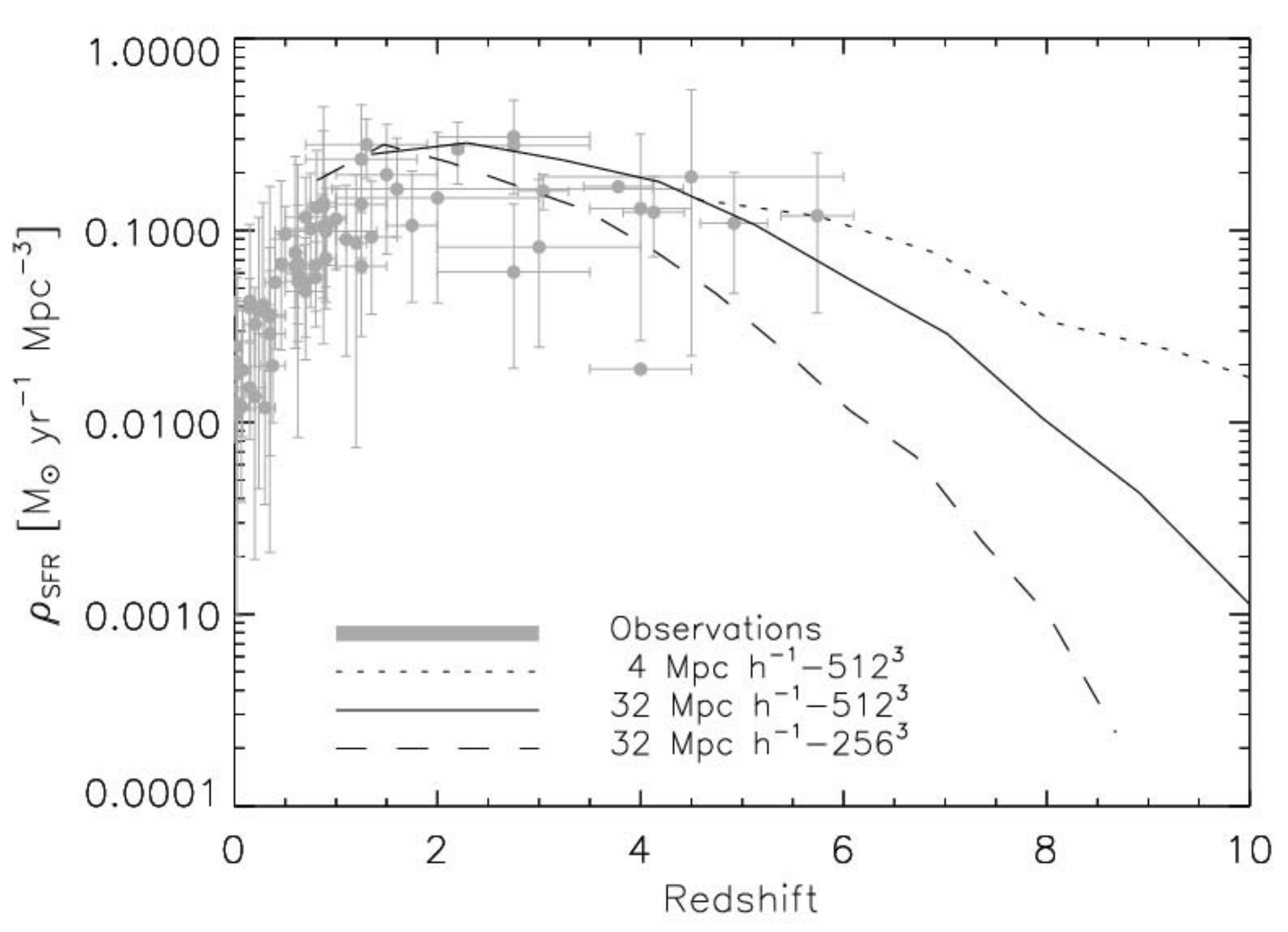}
\caption{Comparison of star formation histories of the universe
between the simulation results (three different lines) and the
observations (shaded symbols with error bars). The simulation results
are differed by resolutions, different box size and particle number.
The observed data is compiled by \citet{hop06}.}
\label{fig10}
\end{figure}
Now, we turn our attention to the application of EU$N$HA to cosmological
star formation.
We solve the equation of motion of the gas and
dark matter particles in a cubic simulation box of a side length
$L_{\rm box}= 32 ~h^{-1}{\rm Mpc}$. The total number of gas and dark
matter particles is $512^3$. The cosmological model adopted in the
simulation is
the WMAP-5 year cosmology with parameters of $\Omega_m=0.26$, $\Omega_{\Lambda}=0.74$,
$\Omega_b=0.044$, $\sigma_8=0.76$, $n_s=0.96$, and  $h = 72~\mathrm{km~
s^{-1}~Mpc^{-1}}$. The simulation starts with glass-like pre-initial condition,
which has a virtually null net gravitational force.
To perturb pre-initial particle distribution, we applied the second-order
linear perturbation method described in \citet{jen10}.
The initial power spectra for baryonic and dark matter at
redshift $z=150$ are calculated by the CAMB package (http://camb.info/sources).
The initial background temperature of gas is calculated using the RecFast package \citep{sea99}.
In generating initial conditions,
the density fluctuations of the gas particles may create temperature fluctuations.
We determine the temperature fluctuation for each gas particle using the adiabatic model as
\begin{equation}
\Delta\ln~T =(\gamma-1)\Delta\ln\rho.
\end{equation}

Figure \ref{fig08} is an example of the gas density at $z=6$. The
distribution of gas particles by and large follow
those of dark matter particles such as clusters, filamentary structures,
and cosmic voids.
After the predefined cosmic reionization epoch at $z_{re}=8.9$,
most of gas particles in low density regions of $n_{\mathrm{H}}<0.014
~\mathrm{cm}^{-3}$ are heated up to $T_g\sim10^4$ K, thus the distribution
of the gas particles is more diffuse than dark
matter particles. Meanwhile, gas particles located in high density regions
($n_{\mathrm{H}}>0.014~\mathrm{cm}^{-3}$), which are optically thick to the
cosmic UV background, may cool down to lower temperature and, thus, settle
down to the inner region of dark matter halos. The detailed
density-temperature relation of gas particles are shown in Figure
\ref{fig09}.
Due to the gas inflow to the halo center, typically
halo regions have $\nabla\cdot\mathbf{v}<0$.
And as
the inner regions of the halo have a high hydrogen density and
low temperature, gas particles are able to cool and finally form stars therein.
The red dots in Figure \ref{fig08}
represent the newly generated stars in clustered regions.

The observed global star formation rate ($\rho_{\mathrm{SFR}}$) of the Universe is
known to be a function
of redshift with a peak located among $z=2$--3. In Figure \ref{fig10},
we compare the star formation history of simulations (lines) with
observations (symbols, \citealt{hop06}). The overall star formation history
agrees with each other.

However, there are differences in the gradient of star formation rate ($\dot{\rho}_{\mathrm{SFR}}$) between three different simulations, and it is because of the different resolution. \citet{cho12} reported that the lower mass galaxies are preferred sites for star formation in the higher redshift, thus slope of $\rho_{\mathrm{SFR}}(z)$ for the finer resolution is shallower than others. Among cosmological simulations, the highest resolution simulation with $L_{\rm box}=4 ~h^{-1}{\rm Mpc}$ and $512^3$ particles show the lowest $\dot{\rho}_{\mathrm{SFR}}$ but highest $\rho_{\mathrm{SFR}}$.

\section{Summary and Conclusion}

We have developed a new cosmological hydrodynamic simulation code
(EU$N$HA) by combining the pre-existing $N$-body code of GOTPM \citep{dub04}
with the standard smoothed particle hydrodynamics (SPH).
This code fully exploits the advantage of the Oct-Sibling Tree (OST) of the GOTPM
to identify the $N$ nearest neighbors.

For astrophysical evolution of gas particles, we also implemented the
processes of (1) non-adiabatic evolution of the gas particles through radiative heating/cooling,
(2) global reionization, (3) star formation, and (4) energy and metallicity feedbacks by
supernova type II explosion. To demonstrate
our new implementations of the SPH and the astrophysical
processes, we made five test simulations: (1) one-dimensional Riemann problems, (2) Kelvin-Helmholtz instability, (3) three-dimensional blast shock wave, (4) star formation on the isolated galactic disk, and (5) global star formation history in the cosmological context.

It is interesting to see the growth of the shear vortex in the Kelvin-Helmholtz instability test
because the only difference in EU$N$HA from other standard SPH methods
is the neighbor searching.
Even though
it is premature to conclude in this simple test,
we may get a hint from \cite{pri08}, who showed that
the particle noise on the shear contact plane may be one of the possible causes to the
suppressed KH instabilities. Our improved neighbor searching method may reduce
this kind of neighboring noise in the predict-correct method.
The SPH measures hydrodynamic quantities of a gas particle
using nearby interacting neighbors contained by a finite smoothing length.
Therefore, the gas density, pressure, and corresponding
acceleration may change for different smoothing lengths. It means that
during a simulation run a sudden change in the number of interacting neighbors
may also have a sudden impact on the gas motion. Then, small perturbations may be buried
in the numerical noise and may be suppressed from developing on the contact layer.
However, it is too hasty to jump to the conclusion with this simple result
because some authors
already showed that with different initial conditions
the KH vortex can grow even in the standard-SPH test \citep{hopkins13}
but with much less surface mixing than shown in other grid-based tests.
Also \cite{pri08} showed that a standard SPH test may yield a small KH vortex
by adjusting the test parameters. So, it is unclear whether our implementation of the
advanced neighbor searching solves the KH instability problem in the Lagrangian code.
Our tentative conclusion is that our code may develope the KH vortex but the surface mixing
on the contact plane seems to be not so strong as other improved versions of the SPH or
Eulerian codes.
More rigorous tests
will be necessary to draw any conclusion on this issue.

The EU$N$HA code is originally intended to serve as
the basis for the further development of the cosmological hydrodynamic simulation code.
Most hydrodynamic routines
are built for fast and efficient communication between the
parallel ranks, and the SPH functions is carefully organized to be isolated (or modularized)
for flexibility in case of future updates.
Therefore, we may be able to easily exchange the SPH routines with those of
another type of the particle-based
hydrodynamic algorithms.


\acknowledgments

Authors thank the anonymous referee for his/her invaluable comments on the draft, which help
us enhance the consistency of the content.
This work was supported by the BK21 plus program through the National Research Foundation (NRF) funded by the Ministry of Education of Korea. SSK and JS's works were supported by Mid-career Research Program (No. 2011-0016898) through the NRF grant funded by the Ministry of Science, ICT and Future Planning of Korea. JS deeply appreciates Jeong-Sun Hwang for her help in making the initial condition of a compound galaxy.


\end{document}